\documentclass[pra,12pt,a4paper,amsmath,amssymb]{revtex4}

\usepackage{color,epsfig,rotating,graphicx,subfigure}
\usepackage{amsmath,amssymb,amscd}

\usepackage[latin1]{inputenc}
\usepackage[english]{babel}

\usepackage{dsfont}
\usepackage{textcomp}

\renewcommand{\Im}{{\rm Im}}

\newcommand{\ri}{{\rm i}}
\newcommand{\re}{{\rm e}}
\newcommand{\rd}{{\rm d}}

\newcommand{\rp}{{\rm p}}

\newcommand{\kb}{K_{\rm B}}

\newcommand{\Tr}{{\rm Tr}}

\begin{document}
%%%%%%%%%%%%%%%%%%%%%%%%%%%%%%%%%%%%%%%%%%%%%%%%%%%%%%%%%%%%%%%%%%%%%%%%%%%%%%%%%%%
%
% Title Page
%
%%%%%%%%%%%%%%%%%%%%%%%%%%%%%%%%%%%%%%%%%%%%%%%%%%%%%%%%%%%%%%%%%%%%%%%%%%%%%%%%%%%
\title{Dynamics of heat transfer between nano systems}

\author{S.-A. Biehs}

\affiliation{Institut f\"{u}r Physik, Carl von Ossietzky Universit\"{a}t,
D-26111 Oldenburg, Germany.}

\author{G.\ S.\ Agarwal}

\affiliation{Department of Physics, Oklahoma State University, Stillwater, Oklahoma 74078, USA}

\date{\today}

\pacs{}

\begin{abstract}
We develop a dynamical theory of heat transfer between two nano systems. 
In particular, we consider the resonant heat transfer 
between two nanoparticles due to the coupling of localized surface modes having a finite 
spectral width. We model the coupled nanosystem by two coupled quantum mechanical oscillators, 
each interacting with its own heat bath, and obtain a master equation for the dynamics
of heat transfer. The damping rates in the master equation are related to the lifetimes
of localized plasmons in the nanoparticles. We study the dynamics towards the steady state 
and establish connection with the standard theory of heat transfer in steady state.
For strongly coupled nano particles we predict Rabi oscillations in the mean occupation 
number of surface plasmons in each nano particle.
\end{abstract}

\maketitle
%\tableofcontents
\newpage

%%%%%%%%%%%%%%%%%%%%%%%%%%%%%%%%%%%%%%%%%%%%%%%%%%%%%%%%%%%%%%%%%%%%%%%%%%%%%%%%%%%%%%%%%%%%%%%
%
% Introduction 
%
%%%%%%%%%%%%%%%%%%%%%%%%%%%%%%%%%%%%%%%%%%%%%%%%%%%%%%%%%%%%%%%%%%%%%%%%%%%%%%%%%%%%%%%%%%%%%%%

\section{Introduction}

The theory of heat transfer between nanosystems has been developed for a steady state which is
usually characterized by time scales much larger than the relaxation time scale of surface plasmon
excitations. However, such time scales are nowadays accessible experimentally~\cite{VasaEtAl2012}. In particular,
in the case of long range plasmons~\cite{SaridEtAl} these relaxation times could be quite long, demanding the 
study of the dynamics of heat transfer. 

In addition, nearly all the works studying radiative
heat transfer at the nanoscale rely on Rytov's flucutational elextrodynamics~\cite{RytovEtAl89} which is 
based on the fluctuation-dissipation theorem of the second kind~\cite{EckardtPRA291984,AgarwalPRA111975}, 
and on macroscopic Maxwell equations. This theoretical framework was very succesful in the past. Indeed, 
the seminal works by Lifshitz, and by Polder and van Hove~\cite{E.M.Lifshitz_1956,Polder1971}, paved the 
way for numerous studies of Casimir-Lifshitz forces and nanoscale radiative heat transfer. 
As a matter of fact, the main assumption in Rytov's theory is that the considered systems are in 
local thermal equilibrium. By introducing macroscopic thermal fluctuating currents or dipole moments 
it is therefore possible to relate the correlation functions of the currents or dipoles to the 
material properties by applying the fluctuation-dissipation theorem. Therefore, the theory is a 
phenomenological one which has no microscopic justification and enables one to study stationary
or quasi-stationary situations only~\cite{TschikinEtAl2012}. A genuinely microscopic 
quantum mechanical description for the heat transfer problem was provided by 
Janowicz {\itshape et al.}~\cite{Janowicz2003} using the Caldera-Legget model. But again, this work 
restricts itself to the steady state, showing that this quantum mechanical description leads to the 
same results for the heat transfer problem as Rytov's theory.  

In this work, we will introduce a simple quantum mechanical model which allows us to study the 
dynamics of the heat transfer rate between two nanosystems in general and between two 
nanoparticles in particular. To keep our model simple we will right from the start assume 
that we have only two nanoparticles supporting localized surface modes which are assumed to 
provide the main heat flux channel~\cite{Domingues2005}. By this assumption we neglect any 
contribution due to eddy currents~\cite{ChapuisAPL2008,Tomchuk2006}, crossed electromagnetic interaction~\cite{Abajo2012}, 
multipoles~~\cite{Perez2008}, non-Debye relaxation inside the nanoparticle~\cite{Perez2009}, and many-particle
effects~\cite{PBA2011}. Then, the nanoparticles can be modeled as quantum mechanical oscillators which are coupled due 
to the interaction of the localized surface modes through their electromagnetic fields. Since we 
are interested in the heat flow, both nanoparticles are coupled to their own heat baths. By
deriving the master equation for this system we can determine the dynamics of the heat transfer
rate towards the steady state. We establish the connection of our model to the results obtained from
Rytov's theory~\cite{VolokPersson2001,ChapuisAPL2008,DedkovKyasov2011} by comparison with the
steady state solutions of our model.

The structure of the paper is as follows: In section II we motivate and introduce
our model. The master equation is derived in section III, where we also determine
the analytical solutions as well as the short- and large-time limits of these solutions
and the lowest order perturbation results. In section IV we define the heat transfer
rate and show how it can be derived using Fermi's golden rule. Finally, in section V
we compare the steady-state result of our model with the known steady-state solutions
for the heat transfer rate between two nanoparticles determining the coupling constant
parameter of our model. In this section we also discuss the relaxation dynamics of
the heat transfer rate between two nanoparticles.

%%%%%%%%%%%%%%%%%%%%%%%%%%%%%%%%%%%%%%%%%%%%%%%%%%%%%%%%%%%%%%%%%%%%%%%%%%%%%%%%%%%%%%%%%%%%%%%
%
% Master equation
%
%%%%%%%%%%%%%%%%%%%%%%%%%%%%%%%%%%%%%%%%%%%%%%%%%%%%%%%%%%%%%%%%%%%%%%%%%%%%%%%%%%%%%%%%%%%%%%%

\section{Physical model for the dynamics of heat transfer between nano systems}

The aim of our work is to study the heat transfer dynamics between two nanosystems. 
In particular, we model the resonant heat transfer between two nanoparticles which 
support localized surface modes by means of a quantum mechanical approach. Our model is
motivated by the fact that nanoparticles or nanosystems which have an extension
smaller than the wavelength of the impinging electromagnetic field can be described
by a dipole moment $\mathbf{p}$ which is induced by the field. This induced dipole moment is related
to the incoming field $\mathbf{E}$ by~\cite{Jackson}
\begin{equation}
  \mathbf{p} = \epsilon_0 \alpha(\omega) \mathbf{E}
\end{equation}    
where $\alpha$ is the polarizability of the nanoparticle and $\epsilon_0$ is the
permittivity of vacuum. The resonance of the nanoparticle is determined by the poles
of the polarizability. For metallic nanoparticles this resonance can be attributed
to collective charge density oscillations within the particle which are called 
localized surface plasmons.

Considering a very small spherical nanoparticle the polarizability can
be expressed as~\cite{Jackson}
\begin{equation}
  \alpha(\omega) = 4 \pi r^3_0 \frac{\epsilon(\omega) - 1}{\epsilon(\omega) + 2},
\end{equation}
where $r_0$ is the radius of the nanoparticle and $\epsilon$ is its permittivity.
From this expression for the polarizability it becomes apparent that the resonance of
the nanoparticle is given by $\epsilon(\omega) = -2$. This implicit equation can in general 
be solved for a complex frequency $\omega = \omega_{\rm sp} - \ri \Gamma$ which 
determines the oscillation frequency $\omega_{\rm sp}$ of the localized charge 
oscillation and its damping or spectral width $\Gamma$ due to losses inside the nanoparticle.
For Drude metals with the permittivity
\begin{equation}
  \epsilon(\omega) = \epsilon_{\infty} - \frac{\omega_{\rm p}^2}{\omega (\omega + \ri \gamma)}
\label{Eq:DrudeModel}
\end{equation} 
where $\omega_{\rm p}$ is the plasma frequency, $\epsilon_\infty$ the background permittivity and $\gamma$ a phenomenological 
damping constant we find $\omega_{\rm sp} = \omega_\rp/\sqrt{\epsilon_\infty + 2}$ and $\Gamma = \gamma/2$
for $\omega_\rp \gg \gamma$ which is in general the case.

\begin{figure}[Hhbt]
\epsfig{file = 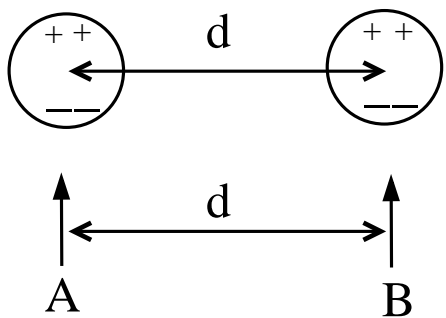, width = 0.35\textwidth}
\caption{\label{Fig:SketchNanoParticle} Sketch of the situation considered. Two nanoparticles separated by a distance $d$
         are effectively replaced by two dipoles oscillating with the surface mode resonance frequency $\omega_{\rm sp}$.}
\end{figure}

Let us now assume that we have two identical nanoparticles $A$ and $B$ as sketched in Fig.~\ref{Fig:SketchNanoParticle}. Particle $A$
has some nonzero temperature $T_1$ and particle $B$ has temperature $T_2  = 0$. Then
localized surface modes will be excited inside particle $A$ due to the 
thermal fluctuations of the charges inside that particle if $\kb T_1 \approx \hbar \omega_{\rm sp}$,
where $\kb$ is Boltzmann's constant and $\hbar$ is Planck's constant. 
The thermally excited resonant charge oscillations generate a dipole field which will 
excite localized charge oscillations inside particle $B$. Since the charge oscillations
are damped inside particle $B$ a part of the exitation energy will be converted into heat and
particle $B$ heats up, whereas particle $A$ cools down. Of course, the charge oscillations
inside particle $B$ will also excite charge oscillations in particle $A$ and so forth. 
So that we can expect that energy will be transferred back and forth between the particles
until a steady state or equilibrium situation is reached.

The heat flux between two nanoparticles within the steady state can be calculated 
by using the fluctuation-dissipation theorem~\cite{EckardtPRA291984,AgarwalPRA111975} and was for example 
studied in Ref.~\cite{ChapuisAPL2008}. Here, we are interested in the dynamics of the 
relaxation towards the steady state which amounts to apply a full non-equilibrium
description. To this end, we first chose to describe the two nanoparticles as 
quantum mechanical oscillators $A$ and $B$ having a frequency $\omega_{\rm sp}$. 
The Hamiltonian of the two oscillators is
\begin{equation}
  H_0 = \hbar \omega_{\rm sp} a^\dagger a + \hbar \omega_{\rm sp} b^\dagger b.
\end{equation}
Here $a^\dagger$, $a$ and $b^\dagger$, $b$ are the creation and annihilation operators
associated to oscillator $A$ and $B$ which fulfill the bosonic commutation relations
$[a,a^\dagger] = [b,b^\dagger] = 1$. Note, that the zero point contribution is
not taken into account, since it does not affect the dynamics.

The interaction of the two nanoparticles which is attributed to the coupled 
charge oscillations is now taken into account by a linear coupling between
oscillator $A$ and $B$ of the form
\begin{equation}
  H_{\rm I} = \hbar g (b^\dagger a + a^\dagger b).
\label{Eq:HamiltonianI}
\end{equation}
That means we allow for the exchange of photons (single photon process) between the two oscillators. 
The first term describes the photon transfer from particle $A$ to particle $B$, i.e.\ 
$| n_{\rm A} , n_{\rm B} \rangle \rightarrow | n_{\rm A} - 1, n_{\rm B} + 1 \rangle$,
whereas the second term describes the photon transfer from particle $B$ to
particle $A$, i.e.\ $| n_{\rm A} , n_{\rm B} \rangle \rightarrow | n_{\rm A} + 1, n_{\rm B} - 1 \rangle$.
The quantity $g$ can be estimated by a variety of methods --- one of which is described in Sec V. 
The most direct and general method is to use an analog of the mode-mode coupling theory as in 
case of waveguides \cite{Saleh}. The parameter $g$ is essentially given by the overlap of the 
plasmon field produced by the nanoparticle $A$ with the plasmon field of the nano particle $B$. 
Note that if $g \ll \Gamma$ then we are in the perturbative regime, however for $g \gg \Gamma$ 
we have non-perturbative regime where Rabi oscillations occur. The situation is somewhat reminiscent 
of cavity QED~\cite{Haroche}.

\begin{figure}
  \epsfig{file = 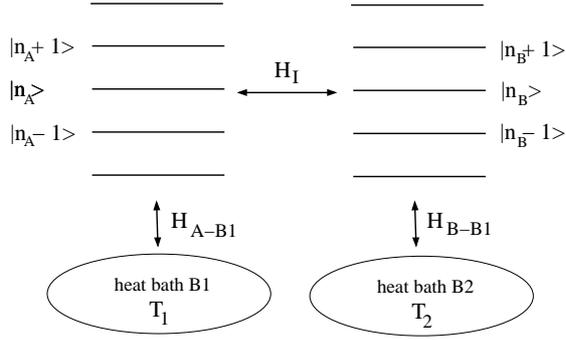, width = 0.45\textwidth}
  \caption{\label{Fig:OscillatorModel} Sketch of our model consisting of two harmonic oscillator $A$ and $B$
           interacting through $H_{\rm I}$. Each oscillator is coupled to a heat bath
           of independend oscillators in equillibrium at temperatures $T_1$ and $T_2$. }
\end{figure}

Since both nanoparticles can attain a temperature we couple each oscillator
to its own heat bath which consist of a spectrum of 
independent oscillators~\cite{AgarwalBook2012} described by the two Hamiltonians
\begin{align}
  H_{B1} &= \sum_j \hbar \omega_{1j} a_{1j}^\dagger a_{1j}, \\
  H_{B2} &= \sum_j \hbar \omega_{2j} a_{2j}^\dagger a_{2j}. 
\end{align}
Again the creation and annihilation operators $a_{1j}^\dagger$, $a_{1j}$
and $a_{2j}^\dagger$, $a_{2j}$ fulfill the bosonic commutation relations
$[a_{1i}, a_{1j}^\dagger] = [a_{2i}, a_{2j}^\dagger] = \delta_{ij}$;
$\omega_{1j}$ and $\omega_{2j}$ are the oscillator frequencies.
We assume that both heat baths are in thermal equilibrium at temperatures
$T_1$ and $T_2$, i.e.\ the density operators of the heat baths are given by 
\begin{equation}
  \rho_{\rm B1/B2} = \frac{\re^{-\beta_{1/2} H_{\rm B1/B2}}}{\Tr\bigl(\re^{-\beta_{1/2} H_{\rm B1/B2}}\bigr)}
\label{Eq:DensityOperatorHB}
\end{equation}
where $\beta_{1/2} = 1/\kb T_{1/2}$ is the inverse temperature.

We further assume that each nanoparticle described by the harmonic 
oscillators $A$ and $B$ is linearly coupled to its heat bath $B1$ and $B2$. 
The corresponding Hamiltonians are given by
\begin{align}
  H_{A-B1} &= \hbar \ri \sum_j g_{1j} (a + a^\dagger) (a_{1j} - a_{1j}^\dagger),  \\
  H_{B-B2} &= \hbar \ri \sum_j g_{2j} (b + b^\dagger) (a_{2j} - a_{2j}^\dagger)  
\end{align}
introducing the coupling strengths $g_{1j}$ and $g_{2j}$ which are related to 
the damping $\Gamma$ of the charge oscillations modeled by the oscillators $A$ and $B$. 
When assuming that both reservoirs are in thermal equilibrium at temperatures $T_1$ and $T_2$ 
and there is no interaction between oscillator $A$ and $B$ ($g = 0$),
then by means of this coupling to their heat baths both oscillators will in the long-time limit
reach an equilibrium state described by the reduced density operators
\begin{equation}
  \rho_{\rm A/B} = \frac{\re^{-\beta_{1/2} H_{\rm A/B}}}{\Tr\bigl(\re^{-\beta_{1/2} H_{\rm A/B}}\bigr)}
\end{equation}
where $H_{\rm A} = \hbar \omega_{\rm sp} a^\dagger a$ and $H_{\rm B} = \hbar \omega_{\rm sp} b^\dagger b$.
Hence, the two harmonic oscillators $A$ and $B$ would acquire the temperatures of their heat baths if
there is no coupling, i.e.\ $g = 0$. Since in our model the coupling $g \neq 0$ between the 
oscillators $A$ and $B$ they will in general not acquire the temperatures of their reservoirs 
in the long-time limit.

Putting all the different terms together the Hamiltonian of our model sketched 
in Fig.~\ref{Fig:OscillatorModel} is given by
\begin{equation}
  H = H_0 + H_{\rm I} + H_{\rm B1} + H_{\rm B2} + H_{\rm A-B1} +  H_{\rm B-B2}.
\end{equation}
For later reference we introduce $H_{\rm S} =  H_0 + H_{\rm I}$ which
is the Hamiltonian of a system of two coupled harmonic oscillors without any
reservoir.

\section{Master equation}

In order to determine the dynamics of the system of two coupled harmonic oscillators
we first derive the master equation of the reduced coupled-oscillator system $H_{\rm S}$. 
Using the standard Born-Markov approximation in combination with the rotating wave
approximation~\cite{AgarwalBook2012} and tracing out the bath variables using the density operators
$\rho_{\rm B1}$ and $\rho_{\rm B2}$ from Eq.~(\ref{Eq:DensityOperatorHB}) we obtain the master equation
\begin{equation}
\begin{split}
  \frac{\partial \rho_S}{\partial t} &= -\ri \omega_a [a^\dagger a, \rho_S] - \ri \omega_b [b^\dagger b, \rho_S] \\
              &\quad - \ri g [a^\dagger b + b^\dagger a, \rho_S] \\
              &\quad - \kappa_1 (\overline{n}_1 + 1)\bigl(a^\dagger a \rho_S - 2 a \rho_S a^\dagger + \rho_S a^\dagger a \bigr) \\
              &\quad - \kappa_1 \overline{n}_1 \bigl( a a^\dagger \rho_S - 2 a^\dagger \rho_S a + \rho_S a a^\dagger \bigr) \\
              &\quad - \kappa_2 (\overline{n}_2 + 1)\bigl(b^\dagger b \rho_S - 2 b \rho_S b^\dagger + \rho_S b^\dagger b \bigr) \\
              &\quad - \kappa_2 \overline{n}_2 \bigl( b b^\dagger \rho_S - 2 b^\dagger \rho_S b + \rho_S b b^\dagger \bigr) 
\end{split}
\end{equation}
where the coupling constants 
\begin{align}
  \kappa_1 &= \pi \sum_j g_{1j}^2 \delta(\omega_{1j} - \omega_a), \\
  \kappa_2 &= \pi \sum_j g_{2j}^2 \delta(\omega_{2j} - \omega_b) 
\end{align}
can in our model be identified as the linewidths $\Gamma$ of the localized surface modes. 
We have also introduced the mean occupation numbers 
\begin{equation}
  \overline{n}_{1/2} = \frac{1}{\re^{\hbar \beta_{1/2}\omega_{a/b}} - 1}
\end{equation}
stemming from the trace over the bath oscillators which are assumed to be in thermal equillibrium
at temperatures $T_1$ and $T_2$. For the sake of generality and later use we have here assumed
that the frequencies $\omega_a$ and $\omega_b$ of the oscillators $A$ and $B$ are different.
In a later stage, we will set these frequencies to $\omega_a = \omega_b = \omega_{\rm sp}$. 

The master equation allows us now to determine the dynamical equations of the mean values
$\langle a^\dagger a \rangle$, $\langle b^\dagger b \rangle$, $\langle a^\dagger b \rangle$ and 
$\langle b^\dagger a \rangle$. By using the commutation relations of the creation and 
annihilation operators for both oscillators we obtain the set of equations
\begin{align}
  \frac{\rd}{\rd t} \langle a^\dagger a\rangle &= -\ri g \bigl(\langle a^\dagger b\rangle - \langle b^\dagger a\rangle  \bigr)
                                                  -2 \kappa_1  \langle a^\dagger a\rangle + 2 \kappa_1 \overline{n}_1, \label{Eq:DiffAdaggerA} \\
  \frac{\rd}{\rd t} \langle b^\dagger b\rangle &= -\ri g \bigl(\langle b^\dagger a\rangle - \langle a^\dagger b\rangle  \bigr)
                                                  -2 \kappa_2  \langle b^\dagger b\rangle + 2 \kappa_2 \overline{n}_2, \\
  \frac{\rd}{\rd t} \langle b^\dagger a\rangle &= \Omega_{ab} \langle b^\dagger a\rangle - \ri g \bigl(\langle b^\dagger b\rangle - \langle a^\dagger a\rangle  \bigr), \label{Eq:DiffBdaggerA} \\
  \frac{\rd}{\rd t} \langle a^\dagger b\rangle &= \Omega_{ba} \langle a^\dagger b\rangle - \ri g \bigl(\langle a^\dagger a\rangle - \langle b^\dagger b\rangle  \bigr) \label{Eq:DiffAdaggerB},
\end{align}
where for the sake of clarity we have introduced the new quantities
\begin{align}
  \Omega_{ab} &= - \ri (\omega_a - \omega_b) - \kappa_1 - \kappa_2, \\
  \Omega_{ba} &= + \ri (\omega_a - \omega_b) - \kappa_1 - \kappa_2.
\end{align}

\subsection{Steady state solutions}

The steady state solutions of the dynamical equations can now be obtained by setting
all time derivatives $\frac{\rd}{\rd t} \langle a^\dagger a\rangle, \frac{\rd}{\rd t} \langle b^\dagger b\rangle$ 
etc.\ equal to zero. Solving the resulting set of equations we obtain
\begin{align}
  \langle a^\dagger a\rangle &= \frac{A (\kappa_1 \overline{n}_1 + \kappa_2 \overline{n}_2) + 2 \kappa_1 \kappa_2 \overline{n}_1}{A (\kappa_1 + \kappa_2) + 2 \kappa_1 \kappa_2}, \label{Eq:AdaggerAStSt} \\
  \langle b^\dagger b\rangle &= \frac{A (\kappa_1 \overline{n}_1 + \kappa_2 \overline{n}_2) + 2 \kappa_1 \kappa_2 \overline{n}_2}{A (\kappa_1 + \kappa_2) + 2 \kappa_1 \kappa_2}, \label{Eq:BdaggerBStSt}\\
	  \langle b^\dagger a\rangle &= \frac{\ri g}{\Omega_{ab}} \frac{2 \kappa_1 \kappa_2 (\overline{n}_2 - \overline{n}_1)}{(\kappa_1 + \kappa_2) A + 2 \kappa_1 \kappa_2}, \\
  \langle a^\dagger b\rangle &=  -\frac{\Omega_{ab}}{\Omega_{ba}} \langle b^\dagger a\rangle,
\end{align}
with
\begin{equation}
  A = - g^2 \frac{\Omega_{ab} + \Omega_{ba}}{\Omega_{ab} \Omega_{ba}} = g^2 \frac{2 (\kappa_1 + \kappa_2)}{(\kappa_1 + \kappa_2)^2 + (\omega_a - \omega_b)^2 }.
\end{equation}

The steady state expressions allow us to check the plausibility of our approach. If we set for example the coupling
between the two oscillators to zero ($g = 0$), then we find $\langle a^\dagger a \rangle = \overline{n}_1$ and
$\langle b^\dagger b \rangle = \overline{n}_2$. That means that the two uncoupled oscillators acquire 
the temperatures of their heat baths, as expected. If we cut off one of the heat baths by 
setting $\kappa_1 = 0$ ($\kappa_2 = 0$), then we find
$\langle a^\dagger a \rangle = \langle b^\dagger b \rangle = \overline{n}_2$ ($\langle a^\dagger a \rangle = \langle b^\dagger b \rangle = \overline{n}_1$) which means that in this case both oscillators take the temperature of the remaining heat bath.
Finally, if we let the coupling between the two oscillators go to infinity ($l \rightarrow \infty$) then
we find
\begin{equation}
  \langle a^\dagger a \rangle = \langle b^\dagger b \rangle = \frac{\kappa_1}{\kappa_1 + \kappa_2} \overline{n}_1 + \frac{\kappa_2}{\kappa_1 + \kappa_2} \overline{n}_2.
\end{equation}
In this case, both oscillators are not in an equilibrium state with one of the reservoirs anymore. Their
mean occupation number is the sum of the equilibrium occupation numbers of both reservoirs weighted by
the relative coupling strength. 

\subsection{Full dynamical solutions}

We will now derive the full dynamical solutions by setting $\omega_a = \omega_b = \omega_{\rm sp}$. 
This corresponds to the situation of a resonant coupling in which we are interested and it simplifies 
our problem, since then $\Omega_{ab} = \Omega_{ba} = - (\kappa_1 + \kappa_2)$. It follows that the 
set of four coupled differential equations can be recasted into a set of 
only 3 coupled differential equations by substracting the last two differential equations (\ref{Eq:DiffBdaggerA})
and (\ref{Eq:DiffAdaggerB}) from each other and by considering 
$\langle b^\dagger a \rangle - \langle a^\dagger b \rangle$ as a new dynamical variable. 
Furthermore we assume that the second heat bath has zero temperature, i.e.\ we set $\overline{n}_2 = 0$. 
Then the 3 coupled differential equations can be stated as
\begin{equation}
  \dot{\mathbf{x}} = \mathds{A}\mathbf{x} + \mathbf{a}
\end{equation}
with $\mathbf{x} = (\langle a^\dagger a \rangle, \langle b^\dagger b \rangle, \langle b^\dagger a \rangle - \langle a^\dagger b \rangle)^t$,
$ \mathbf{a} = (2 \kappa_1 \overline{n}_1, 0, 0)$, and
\begin{equation}
  \mathds{A} = \begin{pmatrix} -2 \kappa_1 & 0 & g \\ 0 & - 2 \kappa_2 & -g \\ 2 g & -2g & - (\kappa_1 + \kappa_2) \end{pmatrix}.
\end{equation}

This set of coupled first order differential equations can be solved by hand in a standard fashion. First we determine the eigenvalues
of $\mathds{A}$ by solving the characteristic equation $\det(\mathds{A} - \lambda \mathds{1}) = 0$. We obtain
the three eigenvalues
\begin{align}
  \lambda_1 &= - (\kappa_1 + \kappa_2), \\
  \lambda_{2,3} &= - (\kappa_1 + \kappa_2) \pm \sqrt{(\kappa_1 - \kappa_2)^2 + 4 g^2}. 
\end{align}
Then we diagonalize matrix $\mathbf{A}$ by introducing the matrix $\mathds{S}$ consisting 
of the eigenvectors of $\mathds{A}$ such that
\begin{equation}
  \mathds{S}^{-1} \mathds{A} \mathds{S} = \begin{pmatrix} \lambda_1 & 0 & 0 \\ 0 & \lambda_2 & 0 \\ 0 & 0 & \lambda_3 \end{pmatrix} \equiv \mathds{B}.
\end{equation}
By means of the matrices $\mathds{S}$ and $\mathds{S}^{-1}$ we decouple the set of coupled
differential equations which translate into
\begin{equation}
  \dot{\mathbf{y}} = \mathds{B}\mathbf{y} + \mathbf{b}.
\end{equation}
Here $\mathbf{y} = \mathds{S}^{-1}\mathbf{x}$ and $\mathbf{b} = \mathds{S}^{-1} \mathbf{a}$. Since
$\mathbf{a} = (2 \kappa_1 \overline{n}_1, 0, 0)^t$, $\mathbf{b}$ equals the first column of $\mathds{S}^{-1}$ times
$2 \kappa_1 \overline{n}_1$. We have now three decoupled inhomogeneous differential equations ($i = 1,2,3$)
\begin{equation}
  \dot{y}_i = \lambda_i y_i + b_i.
\end{equation}
which have the solution
\begin{equation}
  y_i = \re^{\lambda_i t} \eta_i + \bigl( \re^{\lambda_i t} - 1\bigr) \frac{b_i}{\lambda_i}
\end{equation}
where the first term represents the homogeneous solution with a prefactor $\eta_i$ which is
determined by the initial conditions at $t = 0$. The second term represents the inhomogenoeus solution
which vanishes for $t=0$.

From these solutions of the decoupled problem, one can calculate the corresponding solutions
of the coupled set of differential equations by $\mathbf{x} = \mathds{S} \cdot \mathbf{y}$. 
Finally, imposing the initial conditions
\begin{equation}
  \langle a^\dagger a \rangle\bigg|_{t = 0} = \overline{n}_1, \quad
  \langle b^\dagger b \rangle\bigg|_{t = 0} = 0, \quad \text{and} \quad
  (\langle b^\dagger a \rangle - \langle a^\dagger b \rangle ) \bigg|_{t = 0} = 0
\end{equation}
which means that at $t = 0$ oscillator $A$ has the temperature of its heat bath, oscillator $B$ has 
zero temperature (which is the temperature of the second heat bath), and that at $t=0$ the photon transfer
is just turned on, one can determine $\eta_i$. 
The solution for $\mathbf{x}$ fulfilling these initial conditions can be written as
\begin{equation}
 \mathbf{x} = \begin{pmatrix}  
              \sum_i x_i \eta_i \biggl( \re^{\lambda_i t} + \frac{2 \kappa_1}{\lambda_i} \bigl[ \re^{\lambda_i t} - 1 \bigr]\biggr)\\ 
              \sum_i y_i \eta_i \biggl( \re^{\lambda_i t} + \frac{2 \kappa_1}{\lambda_i} \bigl[ \re^{\lambda_i t} - 1 \bigr]\biggr)\\ 
              \sum_i \eta_i \biggl( \re^{\lambda_i t} + \frac{2 \kappa_1}{\lambda_i} \bigl[ \re^{\lambda_i t} - 1 \bigr]\biggr) 
\end{pmatrix}
\label{Eq:AnalyticalSolution}
\end{equation}
where
\begin{equation}
  x_i = \frac{g}{2 \kappa_1 + \lambda_i} \quad\text{and} \quad  y_i = -\frac{g}{2 \kappa_2 + \lambda_i},
\end{equation}
and
\begin{align}
  \eta_1 &= \overline{n}_1 \frac{y_3 - y_2}{N}, \\
  \eta_2 &= \overline{n}_1 \frac{y_1 - y_3}{N}, \\
  \eta_3 &= \overline{n}_1 \frac{y_2 - y_1}{N}, \\
  N &= (x_2 - x_1)(y_1 - y_3) + (x_3 - x_1)(y_2 - y_1).
\end{align}

\begin{figure}[Hhbt]
  \subfigure[\,$\langle a^\dagger a \rangle / \overline{n}_1$]{\epsfig{file = 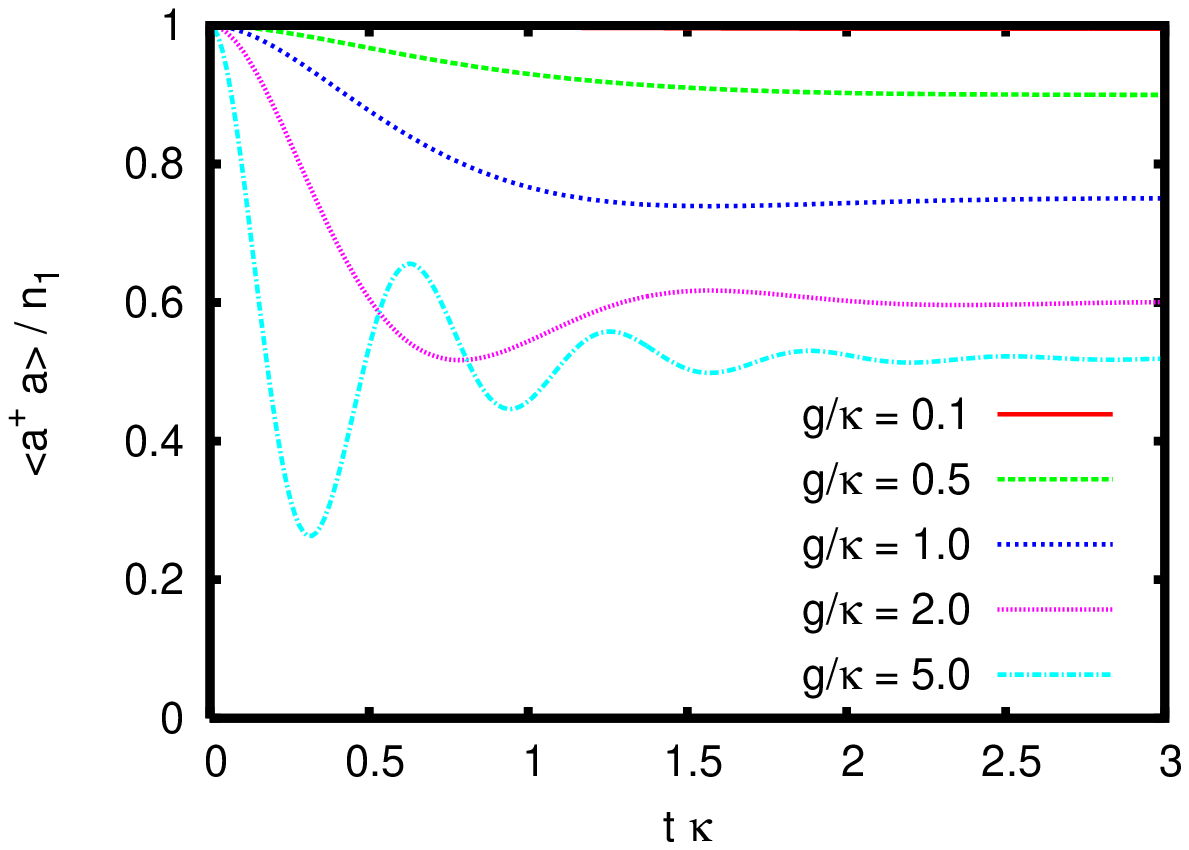, width = 0.45\textwidth}}
  \subfigure[\,$\langle b^\dagger b \rangle / \overline{n}_1$]{\epsfig{file = 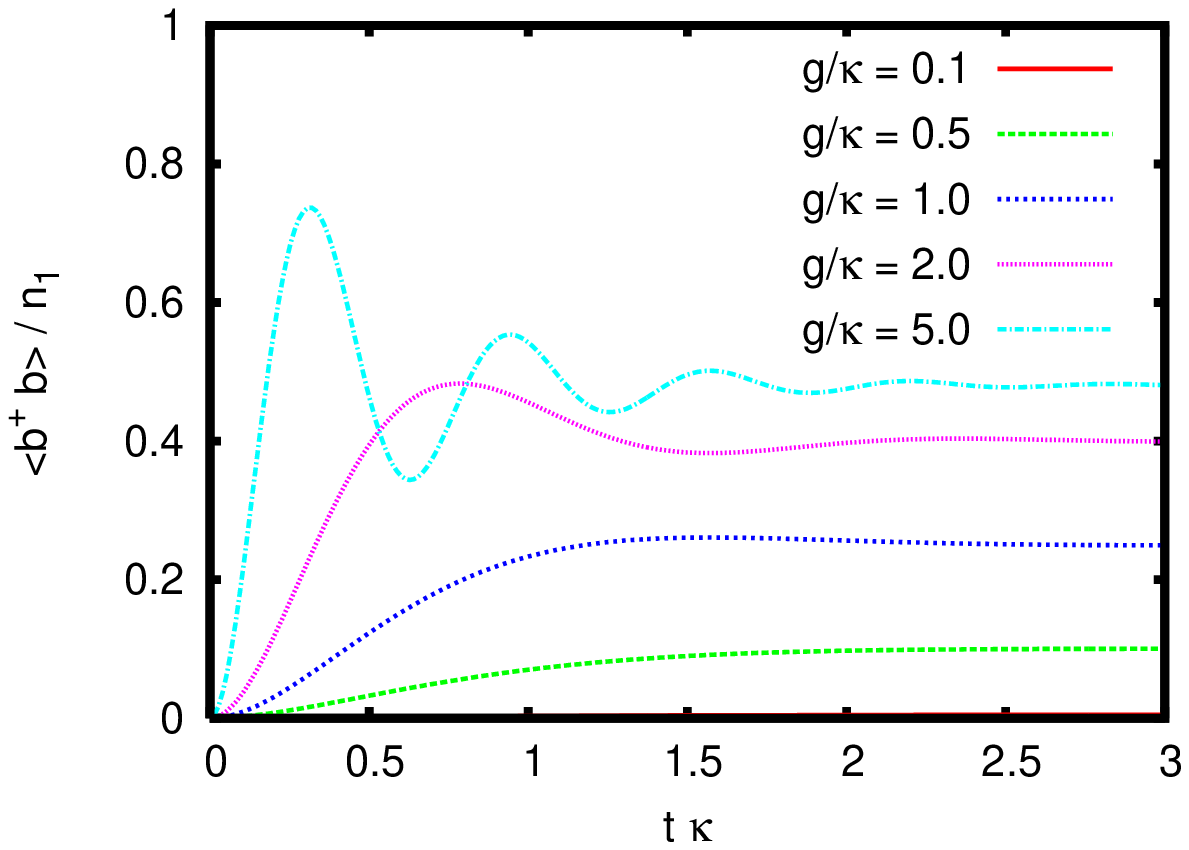, width = 0.45\textwidth}}
  \subfigure[\,$\Im(\langle b^\dagger a \rangle - \langle a^\dagger b \rangle )/\overline{n}_1$]{\epsfig{file = 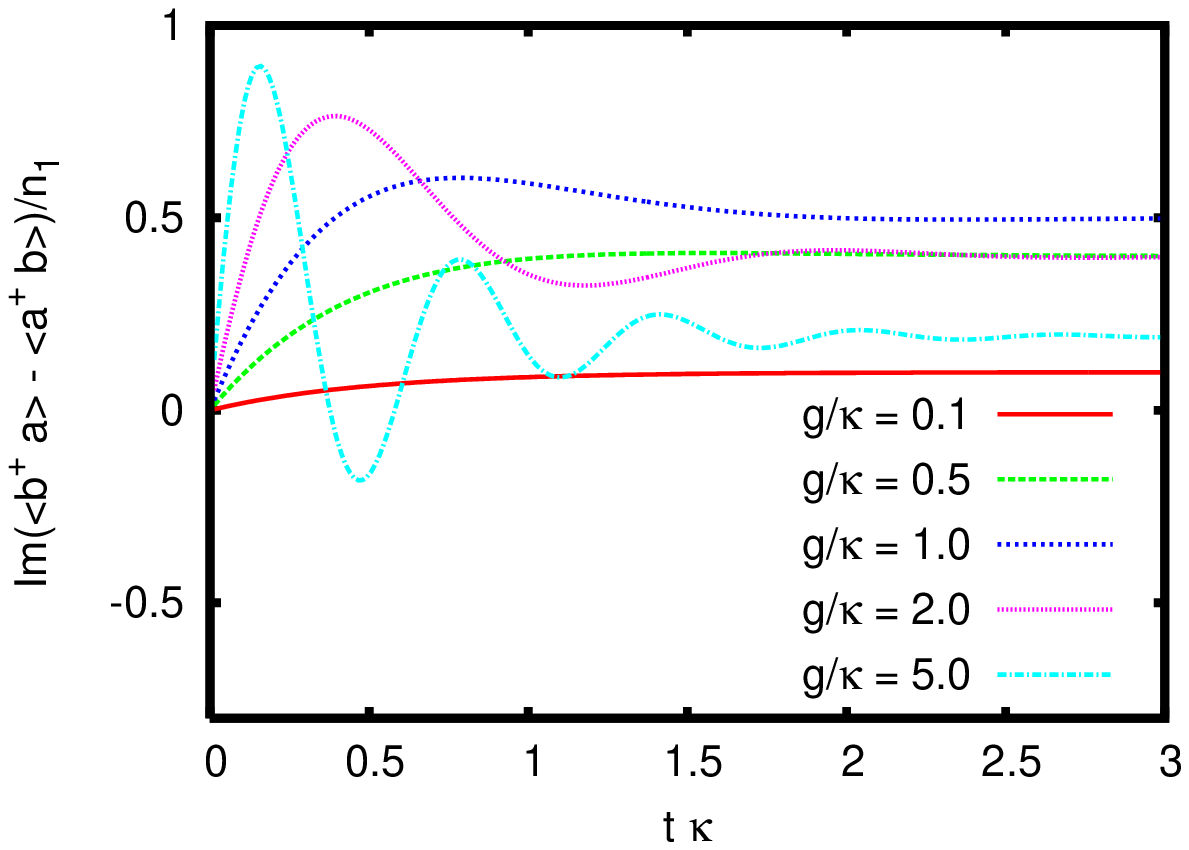, width = 0.45\textwidth}}
  \caption{\label{Fig:AdaggerABdaggerB} Plot of (a) $\langle a^\dagger a \rangle / \overline{n}_1$, (b) $\langle b^\dagger b \rangle / \overline{n}_1$ and (c) $\Im(\langle b^\dagger a \rangle - \langle a^\dagger b \rangle )/\overline{n}_1$ over time for $\kappa_1 = \kappa_2 \equiv \kappa$ and different coupling strengths $g/\kappa$. For large $t$ the solutions converges to the steady state solutions found in Eqs.~(\ref{Eq:AdaggerAStSt}) and (\ref{Eq:BdaggerBStSt}) for $\Omega_{ab} = \Omega_{ba} = - (\kappa_1 + \kappa_2)$ and $\overline{n}_2 = 0$.}
\end{figure}

In Figs.~\ref{Fig:AdaggerABdaggerB} we show some plots of the analytical expressions for $\langle a^\dagger a\rangle /\overline{n}_1$, $\langle b^\dagger b\rangle /\overline{n}_1$, $\Im(\langle b^\dagger a \rangle - \langle a^\dagger b \rangle )/\overline{n}_1$ and  over time choosing $\kappa_1 = \kappa_2 \equiv \kappa$ and different coupling strengths $g/\kappa$ between oscillator $A$ and oscillator $B$. First of all, we can see that the mean occupation number $\langle a^\dagger a \rangle$ of oscillator $A$ reaches in the long-time limit a constant value smaller than its initial value $\overline{n}_1$, whereas the mean occupation number $\langle b^\dagger b \rangle$ of oscillator $B$ converges in the long-time limit to a constant value larger than its initial value $0$. 
Therefore, we observe an energy transfer from oscillator $A$ to oscillator $B$. This energy transfer is related  $\Im(\langle b^\dagger a \rangle - \langle a^\dagger b \rangle )/\overline{n}_1$ as we will see later. 
For small coupling strengths the evolution of the occupation numbers is monotonic, but for large coupling strengths 
one can observe an oscillatory behavior. Hence, the energy is going back and forth between both oscillators until 
the steady state is reached. This behaviour is analogous to the Rabi oscillations observed for 
a two-level system coupled to a field including radiative damping, two coupled qubits, or coupled exciton-plasmon systems for instance~\cite{Loudon,Hakala2009,Tudela2011,VasaEtAl2012}. In such systems Rabi oscillations are found in the strong-coupling limit which 
corresponds in our case to $g / \kappa \gg 1$, whereas in the weak-coupling limit (here  $g / \kappa \ll 1$ ) these oscillations 
are damped out~\cite{Loudon,Tudela2011}.

\begin{figure}[Hhbt]
  \epsfig{file = 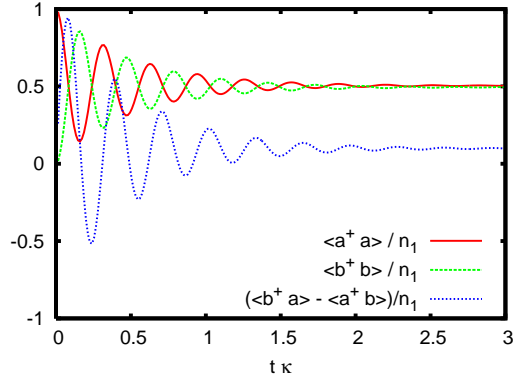, width = 0.45\textwidth}
  \caption{\label{Fig:AdaggerBBdaggerA} Plot of $\Im(\langle b^\dagger a \rangle - \langle a^\dagger b \rangle )/\overline{n}_1$, $\langle a^\dagger a \rangle / \overline{n}_1$, and $\langle b^\dagger b \rangle / \overline{n}_1$ over time for $\kappa_1 = \kappa_2 = 1 \equiv \kappa$ and $g/\kappa = 10$.}
\end{figure}

In Fig.~\ref{Fig:AdaggerBBdaggerA} we show  $\Im(\langle b^\dagger a \rangle - \langle a^\dagger b \rangle )/\overline{n}_1$ over time for $g / \kappa = 10$ together with $\langle a^\dagger a \rangle / \overline{n}_1$ and $\langle b^\dagger b \rangle / \overline{n}_1$.  As we will see later, $\Im(\langle b^\dagger a \rangle - \langle a^\dagger b \rangle )/\overline{n}_1$ is proportional to the energy transfer between oscillator $A$ and $B$. From  Fig.~\ref{Fig:AdaggerBBdaggerA} we see that the oscillations of $\langle b^\dagger b \rangle$ are phase shifted by $\pi$ with respect to the oscillations of $\langle a^\dagger a \rangle$ and the oscillations of $\Im(\langle b^\dagger a \rangle - \langle a^\dagger b \rangle )/\overline{n}_1$ have a phase shift of $\pi/2$ with respect to $\langle a^\dagger a \rangle$ and $\langle b^\dagger b \rangle$. 

Since we are interested in the energy transfer from oscillator $A$ to oscillator $B$ let us further analyse 
$\langle b^\dagger b \rangle$ and $\langle b^\dagger a \rangle - \langle a^\dagger b \rangle$ and derive the
short- and large-time limits. From the analytical solutions we find
\begin{align}
  \langle b^\dagger b \rangle 
        &\approx \begin{cases} 
            \overline{n}_1 g^2 t^2 & t \ll \kappa_1^{-1}, \kappa_2^{-1} \\ 
            \frac{g^2 \overline{n}_1}{\kappa_2 (\kappa_1 + \kappa_2)} \frac{1}{1 + \frac{g^2}{\kappa_1 \kappa_2}}   &   t \gg \kappa_1^{-1}, \kappa_2^{-1} 
        \end{cases} \label{Eq:LimitsAnalyticBdaggerB} \\
  \langle b^\dagger a \rangle - \langle a^\dagger b \rangle 
        &\approx \begin{cases} 
          2 \overline{n}_1 \ri g t & t \ll \kappa_1^{-1}, \kappa_2^{-1} \\ 
          + \frac{2 \ri g \overline{n}_1}{(\kappa_1 + \kappa_2)} \frac{1}{1 + \frac{g^2}{\kappa_1 \kappa_2}}   &   t \gg \kappa_1^{-1},\kappa_2^{-1}
        \end{cases}. \label{Eq:LimitsAnalyticBdaggerA} 
\end{align}
We see that $\langle b^\dagger a \rangle - \langle a^\dagger b \rangle$ is linear in $g t$ and $\langle b^\dagger b \rangle$
is quadratic in $g t$ for times $t \ll \kappa_1^{-1}, \kappa_2^{-1}$. For times $t \gg  \kappa_1^{-1}, \kappa_2^{-1}$ we 
retrieve the steady-state results derived in Eqs.~(\ref{Eq:AdaggerAStSt}) and (\ref{Eq:BdaggerBStSt}) for the case 
$\Omega_{ab} = \Omega_{ba} = - (\kappa_1 + \kappa_2)$ and $\overline{n}_2 = 0$. Furthermore, we see that for $t \gg  \kappa_1^{-1}, \kappa_2^{-1}$
\begin{align}
   \langle b^\dagger b \rangle &\propto g^2 \biggl(1 - \frac{g^2}{\kappa_1 \kappa_2} + \ldots\biggr), \\
   \langle b^\dagger a \rangle - \langle a^\dagger b \rangle &\propto g \biggl(1 - \frac{g^2}{\kappa_1 \kappa_2} + \ldots\biggr).
\end{align} 
From this one can expect that $\langle b^\dagger b \rangle$ has only even orders in $g$, whereas $ \langle b^\dagger a \rangle - \langle a^\dagger b \rangle$ has only odd orders in $g$.

\subsection{Perturbation expansion}

To complete the analysis, we also derive the perturbation result of the set of differential equations (\ref{Eq:DiffAdaggerA})-(\ref{Eq:DiffAdaggerB})
for $\langle b^\dagger b \rangle$ and $\langle b^\dagger a \rangle - \langle a^\dagger b \rangle$ in orders of $g$. 
Here again we first consider the general case $\Omega_{ab} \neq \Omega_{ba}$.  
Expanding $\langle a^\dagger a \rangle = \langle a^\dagger a \rangle^{(0)} + \langle a^\dagger a \rangle^{(1)}  + \langle a^\dagger a \rangle^{(2)} + ... $, $\langle b^\dagger b \rangle = \langle b^\dagger b \rangle^{(0)} + \langle b^\dagger b \rangle^{(1)}  + \langle b^\dagger b \rangle^{(2)} + ... $ etc.\ and solving the dynamical equations for the different perturbation orders.
We find for the lowest orders giving nonvanishing contributions
\begin{equation}
  \langle b^\dagger a \rangle^{(1)} - \langle a^\dagger b \rangle^{(1)} = \ri g \overline{n}_1 \biggl[ \frac{\re^{\Omega_{ab} t} - 1}{\Omega_{ab}} +  \frac{\re^{\Omega_{ba} t} - 1}{\Omega_{ba}} \biggr] 
\end{equation}
and
\begin{equation}
  \langle b^\dagger b \rangle^{(2)} = g^2 \overline{n}_1 \biggl[ \frac{\re^{\Omega_{ab} t} - \re^{-2 \kappa_2 t}}{\Omega_{ab} (2 \kappa_2 + \Omega_{ab})} + \frac{\re^{\Omega_{ba} t} - \re^{-2 \kappa_2 t}}{\Omega_{ba} (2 \kappa_2 + \Omega_{ba})} + \frac{\re^{-2 \kappa_2 t} - 1}{2 \kappa_2} \frac{\Omega_{ab} + \Omega_{ba}}{\Omega_{ab} \Omega_{ba}} \biggr]. \label{Eq:PertResultBdaggerB}
\end{equation}
As expected $\langle b^\dagger b \rangle^{(1)} = 0$.

To compare these results with the analytical solutions we set 
$\omega_a = \omega_b = \omega_{\rm sp}$, i.e.\ $\Omega_{ab} = \Omega_{ba} = - (\kappa_1 + \kappa_2)$. Then we find the short and long time limits
\begin{equation}
 \langle b^\dagger a \rangle^{(1)} - \langle a^\dagger b \rangle^{(1)} 
          \approx \begin{cases} 
           2 \overline{n}_1 \ri g t & t \ll \kappa_1^{-1}, \kappa_2^{-1} \\ 
           \frac{2 \ri g \overline{n}_1}{(\kappa_1 + \kappa_2)}&   t \gg \kappa_1^{-1}, \kappa_2^{-1}
        \end{cases} \label{Eq:LimitsPertBdaggerA}
\end{equation}
and
\begin{equation}
 \langle b^\dagger b \rangle^{(2)} 
          \approx \begin{cases} 
           \overline{n}_1 g^2 t^2 & t   \ll \kappa_1^{-1}, \kappa_2^{-1} \\ 
           \frac{g^2 \overline{n}_1}{(\kappa_1 + \kappa_2) \kappa_2}&   t  \gg \kappa_1^{-1}, \kappa_2^{-1}
        \end{cases} .
\end{equation}
By comparing these perturbation results with the corresponding analytical expressions in (\ref{Eq:LimitsAnalyticBdaggerB}) 
and (\ref{Eq:LimitsAnalyticBdaggerA}) 
we can infer that in general the perturbation results are valid for $g^2 \ll \kappa_1 \kappa_2$. In particular,
we find that for $t \ll \kappa_1^{-1}, \kappa_2^{-1}$ the perturbation result coincides with the analytical solution.

%%%%%%%%%%%%%%%%%%%%%%%%%%%%%%%%%%%%%%%%%%%%%%%%%%%%%%%%%%%%%%%%%%%%%%%%%%%%%%%%%%%%%%%%%%%%%%%
%
% Heat transfer rate
%
%%%%%%%%%%%%%%%%%%%%%%%%%%%%%%%%%%%%%%%%%%%%%%%%%%%%%%%%%%%%%%%%%%%%%%%%%%%%%%%%%%%%%%%%%%%%%%%

\section{Heat transfer rate}

Before we apply our model to describe the dynamics of heat transfer between two nanoparticles, we 
will define the energy transfer rate and discuss under which circumstances it can be derived
from Fermi's golden rule.

\subsection{Defining the energy transfer rate}

For two classical harmonic oscillators $A$ and $B$ the transferred energy per unit time from $A$ to $B$ 
can be described by the rate of work done on oscillator $B$ by oscillator $A$
\begin{equation}
  P = k (x_B - x_A) \cdot \dot{x}_B
\label{Eq:DefHeatTransfer}
\end{equation}
where $k$ is the spring constant for the spring between the oscillators and $x_A$ and $x_B$ 
are the displacements of the oscillators from their equilibrium position. The quantum mechanical analog
can be derived from that expression by replacing
\begin{align}
  x_A &\rightarrow (a^\dagger + a) \sqrt{\frac{\hbar}{2 m \omega_{\rm sp}}}, \label{Eq:DisplacementA} \\
  x_B &\rightarrow (b^\dagger + b) \sqrt{\frac{\hbar}{2 m \omega_{\rm sp}}}, \label{Eq:DisplacementB}\\
  \dot{x}_A = \frac{p_A}{m} &\rightarrow \ri (a^\dagger - a) \sqrt{\frac{\hbar \omega_{\rm sp}}{2 m}}, \\
  \dot{x}_B = \frac{p_B}{m} &\rightarrow \ri (b^\dagger - b) \sqrt{\frac{\hbar \omega_{\rm sp}}{2 m}}.
\end{align}
Keeping only terms for which a photon is interchanged between both oscillators, i.e.\ which
allow for the transition processes $| \overline{n}_1, \overline{n}_2 \rangle \rightarrow | \overline{n}_1 -1, \overline{n}_2 + 1\rangle $
and $| \overline{n}_1, \overline{n}_2 \rangle \rightarrow | \overline{n}_1 + 1, \overline{n}_2 - 1\rangle $, we obtain
\begin{equation}
  P = k \ri \frac{\hbar}{2 m} (a b^\dagger - b a^\dagger).
\end{equation}
In order to relate the prefactor to the coupling constant $g$ or $l$ used in our model,
we start with the classical expression for the potential energy 
\begin{equation}
  H_{\rm I} = \frac{k}{2} (x_A - x_B)^2
\end{equation}
and replace the displacements by the expressions in Eqs.~(\ref{Eq:DisplacementA}) and (\ref{Eq:DisplacementB}).
Keeping again only terms allowing for photon exchange between the oscillators we find
\begin{equation}
  H_{\rm I} = - k \frac{\hbar}{2 m \omega_{\rm sp}} (a^\dagger b + b^\dagger a).
\end{equation}
By comparing this result with the interaction Hamiltonian of our model in Eq.~(\ref{Eq:HamiltonianI})
we find that $g = - k / 2 m \omega_{\rm sp}$. Hence for the transferred power 
\begin{equation}
  P =  \hbar \omega_{\rm sp} (-\ri g) (a b^\dagger - b a^\dagger).
\end{equation}

\subsubsection{Full expression}

Since we know the analytical solution for $\langle a b^\dagger \rangle - \langle a^\dagger b \rangle$ we can immediately
write down the analytical solution for the mean rate of energy transfer
\begin{equation}
  \langle P \rangle = \hbar \omega_{\rm sp} (-\ri g) \bigl[ \langle a b^\dagger \rangle - \langle a^\dagger b \rangle \bigr] = \hbar \omega_{\rm sp} R  
\end{equation}
where the energy transfer rate $R$ is by means of Eq.~(\ref{Eq:AnalyticalSolution}) given by 
\begin{equation}
 R = -\ri g \sum_i \eta_i \biggl[ \re^{\lambda_i t} + \frac{2 \kappa_1}{\lambda_i} \biggl( \re^{\lambda_i t} - 1 \biggr)\biggr].
\label{Eq:RateFullExpression}
\end{equation}
From this expression we can derive the short-time and large-time limits
\begin{equation}
  R \approx \begin{cases} 
          2 \overline{n}_1 g^2 t & t \ll \kappa_1^{-1},\kappa_2^{-1} \\ 
          \frac{2 g^2 \overline{n}_1}{(\kappa_1 + \kappa_2)} \frac{1}{1 + \frac{g^2}{\kappa_1 \kappa_2}} &   t \gg \kappa_1^{-1}, \kappa_2^{-1}
        \end{cases} . 
\label{Eq:RateAnalyticLimits}
\end{equation}	
Note that in the short-time limit the rate $R$ is given by $\partial \langle b^\dagger b \rangle / \partial t$ taking
$\langle b^\dagger b \rangle$ from Eq.~(\ref{Eq:LimitsAnalyticBdaggerB}). On the other hand, in the long-time limit
this observation does not remain true. In the regime of Rabi oscillations ($g \gg \kappa_{1/2}$) the rate of 
transfer is not a meaningful quantity and we have to work with the integrated power.

In the general case ($n_1 > 0$ and $n_2 > 0$) the expression for the transferred power 
in Eq.~(\ref{Eq:DefHeatTransfer}) has to be augmented by a term $- k (x_A - x_B) \cdot \dot{x}_A$ quantifying the 
rate of power done on oscillator $A$ by oscillator $B$. The resulting expression for $P$ and $R$
due to that term is just the same as in Eq.~(\ref{Eq:RateFullExpression}), but with $n_1$ replaced by $-n_2$.
Hence, the overall transferred energy has a prefactor $(n_1 - n_2)$ determining the direction of
the energy transfer or heat flux, uniqely. It follows that in steady state the energy is always transferred from 
the hotter to the colder nanosystem.

\subsubsection{Perturbation result}

The corresponding expression for the heat transfer rate for small coupling strength 
is obtained with the help of Eq.~(\ref{Eq:LimitsPertBdaggerA}), giving
\begin{equation}
 R \approx -\ri g \bigl[ \langle b^\dagger a \rangle^{(1)} - \langle a^\dagger b \rangle^{(1)}  \bigr]
          \approx \begin{cases} 
            2 \overline{n}_1 g^2 t & t \ll \kappa_1^{-1}, \kappa_2^{-1} \\ 
            \frac{2 g^2 \overline{n}_1}{(\kappa_1 + \kappa_2)}& t \gg \kappa_1^{-1}, \kappa_2^{-1}
        \end{cases}. \label{Eq:HeatTransferRatePert}
\end{equation}
As noted above for times $t \ll \kappa_1, \kappa_2$ the perturbation result and the analytical result coincide, whereas
for times $t \gg \kappa_1, \kappa_2$ the perturbation result is only valid for $g^2 \ll \kappa_1 \kappa_2$.

\subsection{Fermi's golden rule}

Finally, we derive the heat transfer rate by using Fermi's golden rule
\begin{equation}
  R_{\rm FGR} = \frac{2 \pi}{\hbar^2} |\langle f | H_{\rm I} | i \rangle |^2 \delta(\omega_f - \omega_i).
\end{equation}
In our case the initial state is given by a) $| i \rangle =  |n_1, n_2 \rangle$ and
$| f \rangle = | n_1 - 1, n_2 + 1 \rangle$
describing the photon transfer from oscillator $A$ to $B$ and 
b) $| i \rangle =  |n_1, n_2 \rangle$ and $| f \rangle = | n_1 + 1, n_2 - 1 \rangle$ 
describing the photon transfer from oscillator $B$ to $A$. Inserting the interaction 
Hamiltonian $H_{\rm I}$ from Eq.~(\ref{Eq:HamiltonianI}) and taking the difference of process a) and b) we find
\begin{equation}
\begin{split}
  R_{\rm FGR} &= 2 \pi g^2 \bigl[ n_1 (n_2 + 1) - n_2 (n_1 + 1) \bigr] \delta(\omega_f - \omega_i) \\
              &= 2 \pi g^2 \bigl[ n_1 - n_2 \bigr] \delta(\omega_f - \omega_i). 
\end{split}
\end{equation}
For the special case that $n_1 = \overline{n}_1$, $n_2 = \overline{n}_2 = 0$, $\omega_f = \omega_b$, and $\omega_i = \omega_a$ we obtain
\begin{equation}
  R_{\rm FGR} = 2 \pi g^2 \overline{n}_1 \delta(\omega_a - \omega_b).
\end{equation}
Now, let us take into account that the oscillators $A$ and $B$ have a finite
linewidth $\kappa_1$ and $\kappa_2$ through their coupling to the reservoirs.
Assuming a Lorentzian profile for the linewidth and averaging over these
profiles, we obtain (setting $\omega_a = \omega_b = \omega_{\rm sp}$) 
\begin{equation}
\begin{split}
  R_{\rm FGR} &=  g^2 \overline{n}_1 2 \pi \int \!\!\rd \omega_1 \, \int\!\!\rd \omega_2\, \frac{\kappa_1/\pi}{(\omega_{\rm sp} - \omega_1)^2 + \kappa_1^2}  \frac{\kappa_2/\pi}{(\omega_{\rm sp} - \omega_2)^2 + \kappa_2^2} \delta(\omega_1 - \omega_2) \\ 
    &= \frac{2 g^2 \overline{n}_1}{\kappa_1 + \kappa_2}.
\end{split}
\end{equation} 
Comparing this expression with the perturbation result for the heat 
transfer rate in Eq.~(\ref{Eq:HeatTransferRatePert}) we see that it is valid for $t \gg \kappa_1, \kappa_2$ and
$g^2 \ll \kappa_1 \kappa_2$.

To have a better understanding of the Fermi golden rule expression for the heat transfer rate, let us 
see how the golden rule is derived from $\langle b^\dagger b\rangle$. First, the golden rule is based
on time-dependent first order perturbation theory in the transition amplitudes. Therefore, for $\langle b^\dagger b \rangle$
we have to make a second-order perturbation expansion with respect to the coupling strength. The resulting 
expression for $\langle b^\dagger b \rangle$ is already given in Eq.~(\ref{Eq:PertResultBdaggerB}). If
we cut off the heat baths by setting $\kappa_1 = \kappa_2 = 0$, then we arrive at
\begin{equation}
  \langle b^\dagger b \rangle^{(2)} = g^2 \overline{n}_1 \frac{\sin^2\biggl(\frac{\Delta t}{2}\biggr)}{\biggl( \frac{\Delta}{2} \biggr)^2},
\end{equation}
where $\Delta = \omega_a - \omega_b$.
From such expression Fermi's golden rule is usually derived by considering the long-time limit~\cite{Loudon}
\begin{equation}
  R_{\rm FGR} = \lim_{t \rightarrow \infty} \frac{\langle b^\dagger b \rangle}{t} =  g^2 \overline{n}_1 2 \pi \delta(\Delta) 
\end{equation}
which is Fermi's golden rule; we have used that
\begin{equation}
  \lim_{t \rightarrow \infty}  \frac{\sin^2\biggl(\frac{\Delta t}{2}\biggr)}{\biggl( \frac{\Delta}{2} \biggr)^2} = 2 \pi t \delta(\Delta).
\end{equation}
Hence, the heat transfer rate can be derived from $\langle b^\dagger b \rangle$ or Fermi's golden rule when first cutting off
the heat baths first. The resulting rate is only valid in the long-time limit for $t \gg \Delta^{-1}$ and for small coupling 
strength, since it is based on perturbation theory. The effect of the coupling to the heat baths is taken into account
by averaging the rate $R_{\rm FGR}$ over the linewidth profiles. For two nanoparticles with $\omega_a = \omega_b = \omega_{\rm sp}$ having
a finite linewidth $\kappa_1$ and $\kappa_2$ the long-time limit condition then translates into $t \gg \kappa_1^{-1}, \kappa_2^{-1}$.

%%%%%%%%%%%%%%%%%%%%%%%%%%%%%%%%%%%%%%%%%%%%%%%%%%%%%%%%%%%%%%%%%%%%%%%%%%%%%%%%%%%%%%%%%%%%%%%
%
% Heat transfer between two Nanoparticles
%
%%%%%%%%%%%%%%%%%%%%%%%%%%%%%%%%%%%%%%%%%%%%%%%%%%%%%%%%%%%%%%%%%%%%%%%%%%%%%%%%%%%%%%%%%%%%%%%

\section{Heat exchange between two nanoparticles - Determination of the parameter $g$ in the microscopic interaction Hamiltonian $H_I$}

Up to here we have discussed our model itself in some detail. To relate it to the heat transfer
dynamics for the resonant radiative heat exchange between two nanoparticles, we still need to 
determine the coupling $g$ for describing the heat transfer rate. To do this, we consider
the heat flux between two nano-particles in the steady-state and determine $g$ from that
by comparing the heat transfer rate with the steady-state solution of our model. 

\subsection{Heat transfer rate in steady state}

The heat transfer rate between two nano-particles in steady state was derived for example 
in Refs.~\cite{VolokPersson2001,Domingues2005,ChapuisAPL2008}. Its derivation is usually based on the fluctuation-dissipation theorem
of the second kind~\cite{EckardtPRA291984,AgarwalPRA111975} which serves as the basis of Rytov's fluctuational 
electrodynamics~\cite{RytovEtAl89}. Within this framework the power $P$ transferred between two identical nanoparticles
with temperatures $T_1 > 0$ and $T_2 = 0$ 
is~\cite{ChapuisAPL2008}
\begin{equation}
  P = \int_0^\infty \frac{\rd \omega}{4 \pi^3} \hbar \omega \overline{n}_1 \Im(\alpha)^2 
       \frac{\omega^6}{c^6} \biggl[ \frac{3}{\bigl(\frac{\omega}{c} d\bigr)^6} + \frac{1}{\bigl(\frac{\omega}{c} d\bigr)^4} + \frac{1}{\bigl(\frac{\omega}{c} d\bigr)^2} \biggr].   
\label{Eq:HeatFluxNanoParticles}
\end{equation}
This expression is only valid for particles with a radius $r_0$ smaller than the dominant thermal wavelength
and an interparticle distance $d$ larger than their radii. Furthermore, multiple-interactions are neglected as well as the
contribution due to eddy currents. 

Now, since we are interested in the power transferred by the resonant
interaction of the localized surface modes of the nanoparticles we expect that the main contribution comes
from the resonance of the polarizability $\alpha$ at frequency $\omega_{\rm sp}$ which allows for approximating
the above expression by
\begin{equation}
  P \approx \hbar \omega_{\rm sp} \overline{n}_1 F(\omega_{\rm sp}) \int_0^\infty \frac{\rd \omega}{4 \pi^3} \Im(\alpha)^2 
\end{equation}
introducing the function
\begin{equation}
  F(\omega) = \frac{\omega^6}{c^6} \biggl[ \frac{3}{\bigl(\frac{\omega}{c} d\bigr)^6} + \frac{1}{\bigl(\frac{\omega}{c} d\bigr)^4} + \frac{1}{\bigl(\frac{\omega}{c} d\bigr)^2} \biggr].
\end{equation}
Therefore the heat transfer rate in steady state is
\begin{equation}
\begin{split}
  R_{\rm st-st} &= \overline{n}_1 F(\omega_{\rm sp}) \int_0^\infty \frac{\rd \omega}{4 \pi^3} \, \Im(\alpha)^2 \\
                &= \overline{n}_1 F(\omega_{\rm sp}) (4 \pi r^3_0)^2 \frac{1}{2} \int_{-\infty}^{+\infty}\frac{\rd \omega}{4 \pi^3}\, \biggl( \frac{3 \Im(\epsilon)}{|\epsilon + 2|^2} \biggr)^2 .
\end{split}
\end{equation}

To further proceed we assume that we have metallic nanoparticles which can be described by the Drude model in Eq.~(\ref{Eq:DrudeModel}).
Inserting this expression in the integrand of $R_{\rm st-st}$ we obtain
\begin{equation}
  R_{\rm st-st} = \overline{n}_1 F(\omega_{\rm sp}) (4 \pi r^3_0)^2 \frac{\omega_{\rm sp}^6\Gamma^2}{2 \pi^3} \biggl(\frac{3}{\epsilon_\infty + 2}\biggr)^2 \int_{-\infty}^{+\infty} \rd \omega \, \frac{1}{(\omega^2 - \omega^2_{\rm sp} + 2 \ri \omega \Gamma)^2 (\omega^2 - \omega^2_{\rm sp} - 2 \ri \omega \Gamma)^2}
\end{equation}
introducing the surface mode frequency $\omega_{\rm sp} = \omega_{\rm p}/\sqrt{\epsilon_\infty + 2}$ and its damping $\Gamma = \gamma/2$.
The frequency integrand has four second-order poles in the complex frequency plane. We compute the integral by taking
the two poles in the upper half plane into account and obtain assuming that $\omega_{\rm sp} \gg \Gamma$ 
\begin{equation}
  R_{\rm st-st} = \overline{n}_1 \frac{\omega_{\rm sp}^2}{\Gamma} \frac{r^6_0}{2} F(\omega_{\rm sp}) \biggl(\frac{3}{\epsilon_\infty + 2}\biggr)^2.
\label{Eq:RateNanoPartStSt}
\end{equation}

\subsection{Comparison with master equation approach}

Since we have the steady-state result for the heat transfer rate between two identical nanoparticles in Eq.~(\ref{Eq:RateNanoPartStSt})
we can determine the coupling constant $g$ by comparing it to the steady-state solution of
our model. Since expression (\ref{Eq:HeatFluxNanoParticles}) does not include multiple interactions 
between the nanoparticles it is in fact a first order perturbation result in $\alpha^2$. Therefore, we compare it
with the perturbation expression for the steady-state solution in Eq.~(\ref{Eq:HeatTransferRatePert}) 
setting $\kappa_1 = \kappa_2 = \Gamma$
\begin{equation}
  R = \frac{g^2}{\Gamma} \overline{n}_1. 
\end{equation}  
By comparison to $R_{\rm st-st}$ we can identify
\begin{equation}
  g =  \omega_{\rm sp} \frac{r^3_0}{\sqrt{2}} \sqrt{F(\omega_{\rm sp})} \frac{3}{\epsilon_\infty + 2}.
\label{Eq:CouplingNanoPart}
\end{equation}
Since to leading order for small distances $F(\omega_{\rm sp}) \propto 1/d^6$ we have $g \propto 1/d^3$
which is typical for a dipole-dipole interaction. 

\subsection{Heat transfer dynamics for two nanoparticles}

\begin{figure}[Hhbt]
  \subfigure[\,$d = 800\,{\rm nm}, g/\Gamma = 0.6 $]{\epsfig{file = 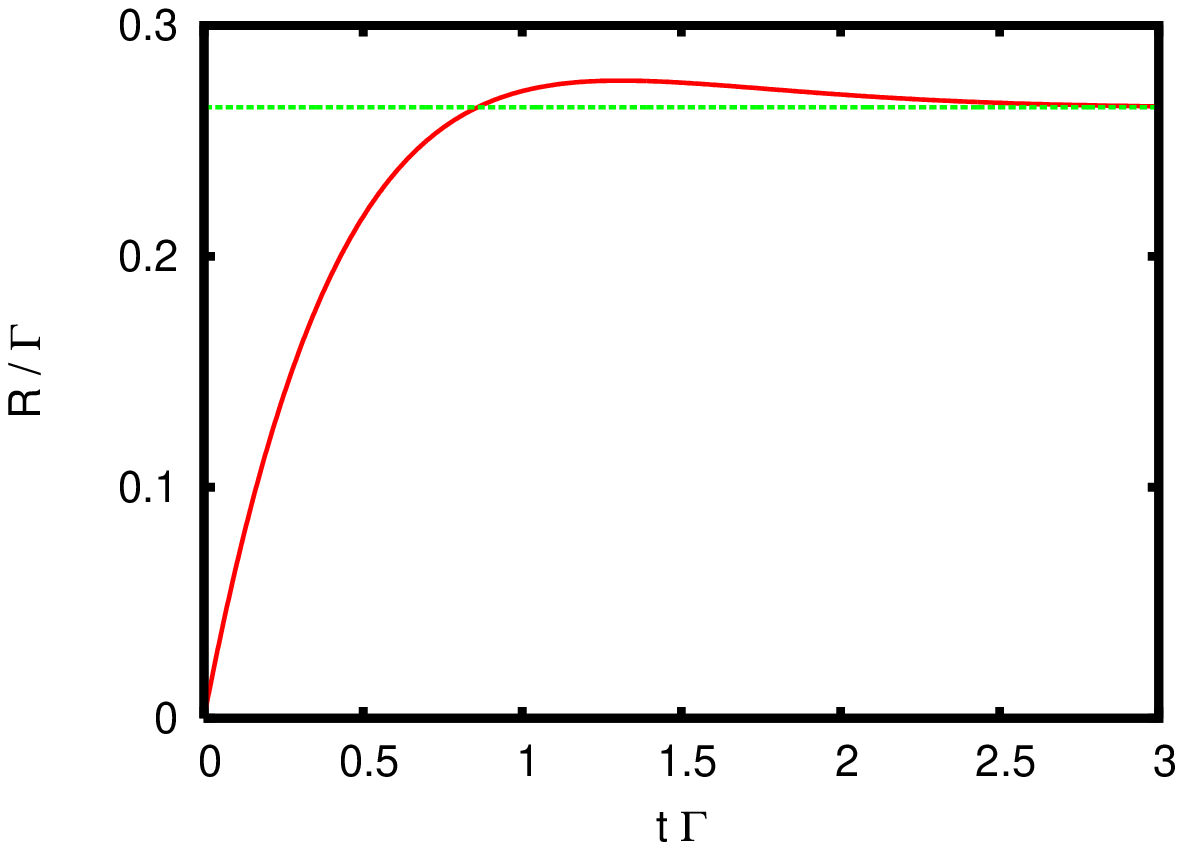, width = 0.45\textwidth}}
  \subfigure[\,$d = 1200\,{\rm nm}, g/\Gamma = 0.4$]{\epsfig{file = 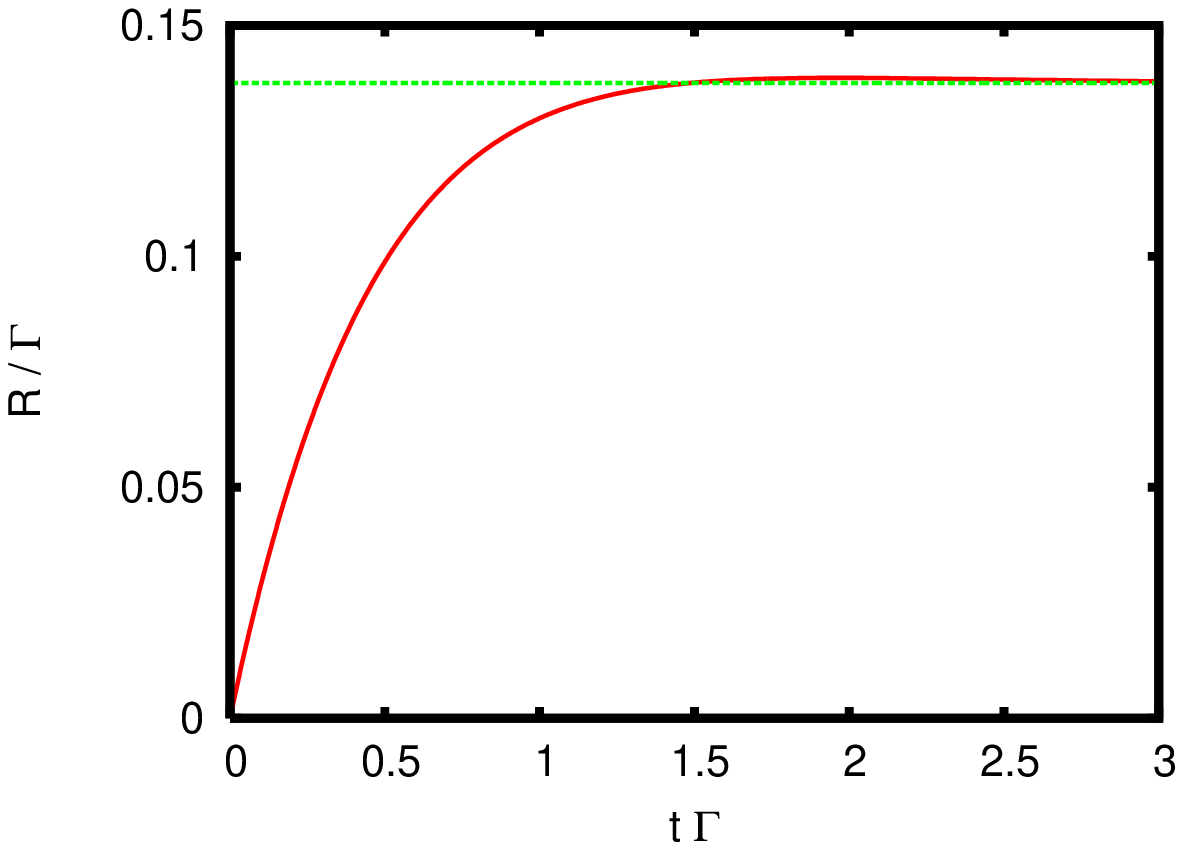, width = 0.45\textwidth}}
  \caption{\label{Fig:HeatTransferRateNanoParticles} Heat transfer rate $R/\Gamma$ from Eq.~(\ref{Eq:RateFullExpression}) between 
           two nanoparticles normalized to the linewidth $\Gamma$ of the nanoparticles surface mode resonances over time in 
           units of $\Gamma^{-1}$. We use the coupling constant $g$ from Eq.~(\ref{Eq:CouplingNanoPart}) with $r_0 = 20\,{\rm nm}$
           and $d = 800\,{\rm nm}$ and $d = 1200\,{\rm nm}$. The horizontal line is the steady-state expression from Eq.~(\ref{Eq:RateAnalyticLimits}).}
\end{figure}

\begin{figure}[Hhbt]
  \epsfig{file = 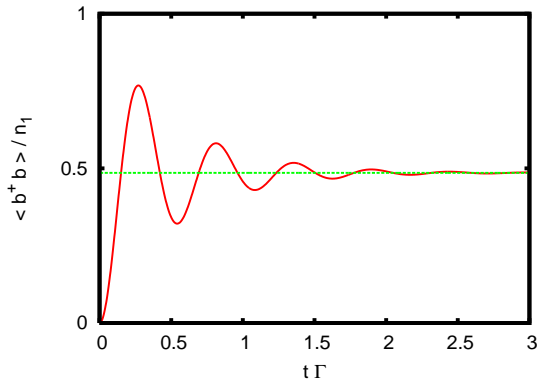, width = 0.45\textwidth}
  \caption{\label{Fig:BdaggerBNanoPart} Mean occupation number $\langle b^\dagger b \rangle / \overline{n}_1$ 
           of the surface mode of nanoparticle $B$ from Eq.~(\ref{Eq:AnalyticalSolution}) over time in 
           units of $\Gamma^{-1}$. We use the coupling constant $g$ from Eq.~(\ref{Eq:CouplingNanoPart}) with $r_0 = 20\,{\rm nm}$
           and $d = 100\,{\rm nm}$, hence $ g/\Gamma = 5.8$. The horizontal line is the steady-state expression from Eq.~(\ref{Eq:LimitsAnalyticBdaggerB}).}
\end{figure}

Finally, we have all the ingredients to determine the dynamics of the heat transfer rate
for two nanoparticles supporting surface modes. When assuming that we have two gold nanoparticles
then $\omega_{\rm p} = 1.4 \times 10^{16} \,{\rm rad}/{\rm s}$, $\epsilon_\infty = 3.7$ and 
$\gamma = 2.79\times10^{13}\,{\rm s}^{-1}$~(see Fig.~2.1 on p.~25 of Ref.~\cite{Soennichsen} for a fit
to the Christy-Johnson data from Ref.~\cite{ChristyJohnson}).
It follows that the surface mode resonance frequency is $\omega_{\rm sp} = 5.86\times10^{15}\,{\rm rad}/ {\rm s}$
with the linewidth $\Gamma = 1.4\times10^{13} \,{\rm s}^{-1}$. If we consider a nanoparticle radius of
$r_0 = 20\,{\rm nm}$ and an interparticle distance of $d = 800\,{\rm nm}$, and $d = 1200\,{\rm nm}$ we find 
the heat transfer rate $R/\Gamma$ plotted in Fig.~\ref{Fig:HeatTransferRateNanoParticles}. 
For times larger than $1/\Gamma$ the heat transfer rate coincides with its steady-state expression.
On the other hand, in the transient regime one can observe the dynamics of the heat transfer rate
for times smaller than $1/\Gamma \approx 72 {\rm fs}$. As can be seen, for the considered distances 
we are mainly in the weak-coupling regime for which we have derived the coupling constant from the known
steady-state heat flux expression. For distances smaller than $d = 800\,{\rm nm}$ the coupling
becomes stronger. Finally, in the strong coupling regime the energy transfer rate is not 
a meaningful quantity so that we consider for this regime $\langle b^\dagger b \rangle /\overline{n}_1$
rather than $R/\Gamma$. In Fig.~\ref{Fig:BdaggerBNanoPart} we plot $\langle b^\dagger b \rangle /\overline{n}_1$
for $d = 100\,{\rm nm}$ which corresponds to $g/\Gamma = 5.8$ showing clearly the Rabi oscillations.

%%%%%%%%%%%%%%%%%%%%%%%%%%%%%%%%%%%%%%%%%%%%%%%%%%%%%%%%%%%%%%%%%%%%%%%%%%%%%%%%%%%%%%%%%%%%%%%
%
% Conclusion
%
%%%%%%%%%%%%%%%%%%%%%%%%%%%%%%%%%%%%%%%%%%%%%%%%%%%%%%%%%%%%%%%%%%%%%%%%%%%%%%%%%%%%%%%%%%%%%%%

\section{Conclusion}

We have introduced a model of two coupled dipoles which are both coupled to their own heat
baths of independent oscillators. This model allows us to describe the heat transfer dynamics between two
nano systems. In particular, we have used this model to determine the heat transfer rate between two
spherical nanoparticles which is due to the resonant interaction of the localized surface modes.
Within our model we have derived the expression in the limit of short times as well as in the 
steady state regime and made a connection to the usual heat transfer calculations which are based
on the fluctuation-dissipation theorem considering the steady-state regime only. 
We have shown that for small coupling strength between the nanoparticles the steady-state heat 
transfer rate can also be determined by Fermi's golden rule when taking into account the finite 
widths of the resonant surface modes. Finally, we have studied the dynamical heat transfer rate for
two gold nanoparticles in the transient regime for times shorter than the 
relaxation time of the surface modes and predict Rabi oscillations for the mean occupation number
of the surface plasmons~\cite{VasaEtAl2012}.

\begin{acknowledgements}
  G.S.A.\ thanks the hospitality of the director of the {\itshape Tata Institute of Fundamental Research}
  in Mumbai where part of this work was done.
\end{acknowledgements}

\end{document}